\documentclass{article}
\usepackage{amsmath}

\begin{document}

\title{Dirac particle in the presence of plane wave and constant magnetic fields: Path integral approach}
\author{S. Bourouaine \thanks{
phybou@wissal.dz} \\
%EndAName
Physics Department, Faculty of Sciences, Mentouri University, \\
Constantine, Algeria} \maketitle

\begin{abstract}
The Green function (GF) related to the problem of a Dirac particle
interacting with a plane wave and constant magnetic fields is calculated in
the framework of path integral via Alexandrou et al. formalism according to
the so-called global projection. As a tool of calculation, we introduce two
identities\ (constraints) into this formalism, their main role is the
reduction of integral's dimension and the emergence in a natural way of some
classical paths, and due to the existence of constant electromagnetic field,
we have used the technique of fluctuations. Hence the calculation of the
(GF) is reduced to a known gaussian integral plus a contribution of the
effective classical action.
\end{abstract}

\maketitle

%aaaaaaa \\




\section{Introduction:\newline
}

The propagator of Dirac particle in an external electromagnetic field is
distinguished from that of the scalar particle by a complicated spin
structure. By using the known anticommuting odd Grassmann variables $\left[ 1%
\right] $, the description of the Dirac propagator gains the possibility to
acquire a representation - Path integral - similar to the case of the scalar
particle modified by a spin factor (SF). This representation was discussed
in various contexts. Nevertheless, the description of Dirac propagator by
only bosonic variables is still unfulfilled. Berezin and Marinov $\left[ 2%
\right] $ showed that massive particles can be described in the usual
five-dimension extension. This idea was exploited by several works, among
them let us quote the successful formalism of Fradkin-Gitman $[3-4]$ in the
relativistic case, in which we note an important supersymmetry between the
bosonic and fermionic parts $\left[ 5\right] $. This formalism saw several
applications while following various computation methods $[6-10].$ Recently,
a generalization of Di Vecchia and Ravndal's approach $\left[ 12-13\right] $
describing a massive Dirac particle in external vector and scalar fields,
and using path integral representations according to the global and local
projections, has been proposed by Alexandrou et al. $\left[ 14\right] $.
This formulation is still endowed with a supersymmetric action which is
derived systematically without inserting a fifth component $\psi ^{5}\left(
\tau \right) $ to the spin variable as opposed to $\left[ 3\right] $.

The purpose of this paper is to develop the problem of a relativistic
particle with mass $m$ in plane wave and constant magnetic fields, and show
the attractive features of spin evolution in the computation of the (SF) by
using the formalism of Alexandrou et al. in its global projection (without
considering the Grassmann proper time) $\left[ 14\right] .$ Note that this
problem has been calculated in the framework of Feynman's approach based on
the T-product $\left[ 15\right] $ which is not strictly a genuine path
integral formulation.

In the first step, by taking into account the definition of the total
potential vector which characterizes our problem, we give the general
formulation of Alexandrou et al.. In the third section, we compute the (GF)
by adopting the fluctuation analysis performed on both, real and Grassmann
variables $\left[ 9\right] $ and inserting the known identities $\left[ 7%
\right] $ into this formulation.

Let us recall initially that the (GF) $S_{g}\left( x_{b},x_{a}\right) $
given by the formalism of Alexandrou et al. in the global projection $\left[
14\right] $ is%
\begin{equation}
S_{g}\left( x_{b},x_{a}\right) =\left( i\gamma ^{\mu }\left( \partial
_{b}-gA_{b}\right) _{\mu }+m\right) G_{g}\left( x_{b},x_{a}\right)
\end{equation}%
with
\begin{align}
G_{g}\left( x_{b},x_{a}\right) & =\frac{-i}{2k_{0}}\exp (i\gamma ^{\mu }%
\frac{\partial }{\partial \Gamma ^{\mu }})\int_{0}^{\infty }dT\int DxDp
\notag \\
& \int_{E}\tilde{D}\psi \exp \left\{ i\int_{0}^{T}\left[ -\frac{k_{0}}{2}%
\dot{x}^{2}-g\dot{x}.A(x)\right. \right.   \notag \\
& +\frac{1}{2k_{0}}\left( p%
%TCIMACRO{\U{b2}}%
%BeginExpansion
{{}^2}%
%EndExpansion
-m^{2}\right) -i\frac{g}{k_{0}}F_{\mu \nu }\psi ^{\mu }\psi ^{\nu }  \notag
\\
& \left. \left. \left. +i\psi .\dot{\psi}\right] d\tau +\psi (0).\psi
(T)\right\} \right\vert _{\Gamma =0,}
\end{align}%
where $g$ is the electron charge, $\gamma^{\mu}a_{\nu}\equiv
\gamma .a$ $\left( \gamma \text{ are the Dirac matrices}\right) $.
The $x,k_{0}$ and $\Gamma $, $\psi $ are respectively the real and
Grassmann (odd) variables. The scalar product of four-vectors,
denoting by a dot, is $a.b=a^{\mu }b_{\mu }.$

The boundary conditions for bosonic variables $x$ are
\begin{equation}
x(0)=x_{a},\quad x(T)=x_{b}
\end{equation}%
and the antiperiodic boundary for the spin variables
\begin{equation}
E=\psi ^{\mu }(T)+\psi ^{\mu }(0)=\Gamma ^{\mu }
\end{equation}%
and we have the proper normalization
\begin{equation}
\tilde{D}\psi =D\psi \lbrack \int_{E}D\psi \exp \{\psi \left( 0\right) .\psi
\left( T\right) -\int_{0}^{T}\psi .\dot{\psi}d\tau \}]^{-1}
\end{equation}

The effective action relative to Eq.$\left( 2\right) $ shows us the
contributions of the spin degrees of freedom to the kinetic energy, as well
as the contribution of coupling the photon to the electron and its spin
evolution.

\section{Formulation of the problem}

We propose a short review of the used notations and conventions in the
definition of electromagnetic field. The total potential vector $A_{\mu
}\left( x\right) $ related to the plane wave and constant magnetic fields is
a sum of two terms%
\begin{equation}
A_{\mu }\left( x\right) =a_{\mu }\left( x^{T}\right) +A_{\mu }^{p}\left(
\varphi \right) ,
\end{equation}%
where $a_{\mu }\left( x^{T}\right) $ generates the constant magnetic field $%
B $ and it is a function of the transverse component of the position vector $%
x$%
\begin{equation}
a_{\mu }\left( x^{T}\right) =\frac{1}{2}fx^{T}=\frac{1}{2}f_{\mu \nu }\left(
x^{T}\right) ^{\nu },\text{ \ }
\end{equation}%
with%
\begin{equation}
f_{\mu \nu }=iB\left( \epsilon _{\mu }\epsilon _{\nu }^{\ast }-\epsilon
_{\nu }\epsilon _{\mu }^{\ast }\right) .
\end{equation}%
$\left( \epsilon ,\epsilon ^{\ast }\right) $ are the basis vectors set, such
as%
\begin{equation}
\epsilon =\frac{1}{\sqrt{2}}\left( 1,i,0,0\right) ,\ \ \ \ \ \ \ \ \epsilon
^{\ast }=\frac{1}{\sqrt{2}}\ \left( 1,-i,0,0\right) \ \ \ \ \ \ \ \ \ \
\end{equation}%
\ satisfying
\begin{equation}
\epsilon .\epsilon =\epsilon ^{\ast }.\epsilon ^{\ast }=0\text{ \ \ and \ \ }%
\epsilon .\epsilon ^{\ast }=1,
\end{equation}%
in which, we can define the transverse components for any vector as
\begin{equation}
\left( x^{T}\right) _{\mu }=\epsilon _{\mu }\left( \epsilon ^{\ast
}.x\right) +\epsilon _{\mu }^{\ast }\left( \epsilon .x\right) .
\end{equation}

$\bigskip A_{\mu }^{p}\left( \varphi \right) $ is the transverse potential
vector of plane wave, and it depends on $\varphi =k.x.$

Notice that the wave vector $k$ has only a longitudinal component
\begin{equation}
k=\left( 0,0,-1,-1\right) \ \ \
\end{equation}%
which implies that $\ k.\epsilon =k.\epsilon ^{\ast }=0,$ with%
\begin{equation}
k.A^{p}\left( \varphi \right) =0\text{ \ \ and \ \ }k%
%TCIMACRO{\U{b2}}%
%BeginExpansion
{{}^2}%
%EndExpansion
=0.
\end{equation}

From $\left( 6\right) ,$ we get the total electromagnetic tensor%
\begin{eqnarray}
F_{\mu \nu }\left( \varphi \right) &=&f_{\mu \nu }+f_{\mu \nu }^{p}\left(
\varphi \right)  \notag \\
&=&iB\left( \epsilon _{\mu }\epsilon _{\nu }^{\ast }-\epsilon _{\nu
}\epsilon _{\mu }^{\ast }\right)  \notag \\
&&+k_{\mu }A_{\nu }^{\prime p}\left( \varphi \right) -k_{\nu }A_{\mu
}^{\prime p}\left( \varphi \right) ,
\end{eqnarray}%
\ where the prime indicates the derivative with respect to $\varphi $.

By making a change on time of integration $\tau \rightarrow \frac{\tau }{%
k_{0}e_{0}}$ and $T=k_{0}e_{0},$ the action of Eq. $\left( 2\right) $ goes
over to that given by the formalism of Fradkin-Gitman (in the limit $\chi
_{0}$(Grassmann proper time) $\rightarrow 0$) with a different sign in the
(SF) and an absence of the fifth Grassmann component, hence%
\begin{eqnarray}
G_{g}\left( x_{b},x_{a}\right) &=&\frac{-i}{2}\exp (i\gamma ^{\mu }\frac{%
\partial }{\partial \Gamma ^{\mu }})\int_{0}^{\infty }de_{0}\int DxDp\int_{E}%
\tilde{D}\psi  \notag \\
&&\exp \left\{ i\int_{0}^{1}\left[ -\frac{\overset{.}{x}^{2}}{2e_{0}}-g\dot{x%
}.A(x)\right. \right.  \notag \\
&&+\frac{e_{0}}{2}\left( p%
%TCIMACRO{\U{b2}}%
%BeginExpansion
{{}^2}%
%EndExpansion
-m^{2}\right) -ie_{0}gF_{\mu \nu }\psi ^{\mu }\psi ^{\nu }  \notag \\
&&\left. \left. \left. +i\psi .\dot{\psi}\right] d\tau +\psi (0).\psi
(1)\right\} \right\vert _{\Gamma =0}
\end{eqnarray}%
with $e_{0}$ is a real variable.

Since the plane wave field is a function of the product $k.x$, it is
preferable to introduce the two following functional identities $\left[ 7%
\right] $%
\begin{gather}
\int d\varphi _{b}d\varphi _{a}\delta \left( \varphi _{a}-k.x_{a}\right)
\int D\varphi Dp_{\varphi }  \notag \\
\exp \left[ i\int_{0}^{1}p_{\varphi }\left( \overset{.}{\varphi }-k.\dot{x}%
\right) d\tau \right] =1
\end{gather}%
and%
\begin{gather}
\int d\eta _{b}d\eta _{a}dp_{\sigma }\int D\eta Dp_{\eta }  \notag \\
\exp \left\{ i\int_{0}^{1}p_{\eta }\left( \dot{\eta}-k.\dot{\psi}\right)
d\tau +ip_{\sigma }\left( \eta _{a}-k.\psi _{a}\right) \right\} =1
\end{gather}%
into expression $\left( 15\right) .$ Because of these identities, the
variables $\varphi $ and $\eta $ are independent respectively of the scalar
products $k.x$ and $k.\psi $.

The term describing the interaction between the spin and a plane wave can be
written as%
\begin{eqnarray}
f_{\mu \nu }^{p}\left( \varphi \right) \psi ^{\mu }\psi ^{\nu } &=&\left[
k_{\mu }A_{\nu }^{\prime p}\left( \varphi \right) -k_{\nu }A_{\mu }^{\prime
p}\left( \varphi \right) \right] \psi ^{\mu }\psi ^{\nu }  \notag \\
&=&2\eta \left( A^{\prime p}.\psi \right) ,
\end{eqnarray}%
in this case, we get%
\begin{align}
G_{g}(x_{b},x_{a})=&\notag
\frac{-i}{2}\exp(i\gamma^{\mu}\frac{\partial}{\partial\Gamma^{\mu}})
\int_{0}^{\infty}de_{0}\int DxDp\int_{E}\widetilde{D}\psi\int
D\eta\int Dp_{\eta}\int d\eta_{b}d\eta_{a}dp_{\sigma}\\\notag&\int
d\varphi_{b}d\varphi_{a}\int
Dp_{\varphi}\delta(\varphi_{a}-k.x_{a})\exp\{i\int_{0}^{1}[-\frac{\dot{x}^{2}}{2e_{0}}-g\dot{x}.A(x)\\\notag
 &+\frac{e_{0}}{2}(p^{2}-m^{2})+p_{\varphi}(\dot{\varphi}-k.\dot{x})-ie_{0}gF_{\mu\nu}\psi^{\mu}\psi^{\nu}
+i\psi.\dot{\psi}-2ie_{0}g\eta
A'^{p}.\psi\\&+p_{\eta}(\dot{\eta}-k.\dot{\psi})]d\tau+\psi(0).\psi(1)+p_{\sigma}(\eta_{a}-k.\psi_{a})\}|_{\Gamma=0},
\end{align}
where $p_{\sigma }$ is a odd Grassmann variable. $\left( \eta
_{b},\varphi _{b},\psi _{b}^{n}\right) $ and $\left( \eta
_{a},\varphi _{a},\psi _{a}^{n}\right) $ are respectively the
variables $\left( \eta \left( \tau \right) ,\varphi \left( \tau
\right) ,\psi ^{n}\left( \tau \right) \right) $ at $\tau =1$ and
$\tau =0$.

\section{Green Function Calculation}

The genuine path integral formulation $\left( 19\right) $ contains all the
dynamics of a Dirac particle moving in the combined field of a plane wave
and a constant magnetic field. By considering the classical trajectories,
the computation of the (GF) is reduced to the computation of the known
gaussian integrals and a contribution of the effective classical action.

Let us use the transverse and longitudinal components of the vectors $\left(
x,p,\psi \right) $ as
\begin{equation}
x=\left(
\begin{array}{c}
x^{T} \\
x^{L}%
\end{array}%
\right) \text{, }p=\left(
\begin{array}{c}
p^{T} \\
p^{L}%
\end{array}%
\right) \text{ \ and \ }\psi =\left(
\begin{array}{c}
\psi ^{T} \\
\psi ^{L}%
\end{array}%
\right) .
\end{equation}%
Therefore, from the definitions $\left( 8\right) ,\left( 11\right) $ and $%
\left( 20\right) $, we get the following scalar products \
\begin{equation}
f_{\mu \nu }\psi ^{\mu }\psi ^{\nu }\equiv \psi ^{T}.\left( f\psi
^{T}\right) ,\ A^{\prime p}.\psi \equiv A^{\prime p}.\psi ^{T}\text{ and }%
k.\psi \equiv k.\psi ^{L}.
\end{equation}%
In order to linearize the quadratic bosonic term along longitudinal plane in
the integral action of Eq. $\left( 19\right) $, and make the vector $p^{L}$
constant during time, we shift%
\begin{equation}
p^{L}\rightarrow p^{L}+\frac{\dot{x}^{L}}{e_{0}}+e_{0}kp_{\varphi }.
\end{equation}

In other words, after considering the transverse and longitudinal components
given by $\left( 20\right) $ and the shifting term $\left( 22\right) ,$ the
successive integrations over $\left( p^{L},x^{L}\right) $ and $\left(
p_{\varphi },\varphi \right) $ then over $p^{T}$ in $\left( 19\right) $,
lead us to

\begin{align}
G_{g}(x_{b},x_{a})=&\notag
\frac{-i}{2}\exp(i\gamma^{\mu}\frac{\partial}{\partial\Gamma^{\mu}})
\int_{0}^{\infty}de_{0}\int\frac{dp^{L}}{(2\pi)^{2}}\int D\eta
Dp_{\eta}\int_{E}\widetilde{D}\psi\int
dp_{\sigma}d\eta_{b}d\eta_{a}
d\varphi_{b}d\varphi_{a}\\\notag&\delta(\varphi_{b}-\varphi_{a}+e_{0}k.p^{L})
\int_{x_{a}^{T}}^{x_{b}^{T}}Dx^{T}\delta(\varphi_{a}-k.x_{a})\exp[ip^{L}.(x_{b}^{L}-x_{a}^{L})\\\notag&+\frac{ie_{0}}{2}(p^{L2}-m^{2})]
\exp\{i\int_{0}^{1}[-\frac{(\dot{x}^{T})^{2}}{2e_{0}}-\frac{g}{2}x^{T}.f\dot{x}^{T}-gA^{p}.\dot{x}^{T}
-ie_{0}g\psi^{T}.f\psi^{T}\\&+i\psi.\dot{\psi}-2ie_{0}g\eta
A'^{p}.\psi^{T}+p_{\eta}(\dot{\eta}-k.\dot{\psi}^{L})]d\tau
+\psi(0).\psi(1)+p_{\sigma}(\eta_{a}-k.\psi^{L}_{a})\}|_{\Gamma=0}
\end{align}
and the extracted scalar path $\dot{\varphi}=-e_{0}k.p^{L}$ with
the evolution
\begin{equation}
\varphi \left( \tau \right) =-e_{0}k.p^{L}\tau +\varphi _{a}.
\end{equation}

Notice that the\ expression $\left( 23\right) $ is obtained after performing
a transformation on vector $p^{T}$ as%
\begin{equation}
p^{T}\rightarrow p^{T}+\frac{\dot{x}^{T}}{e_{0}}+gA\left( x\right) .
\end{equation}

The part which depends on vectors $\left( p^{L},x_{b}^{L},x_{a}^{L}\right) $
in Eq. $\left( 23\right) $ corresponds to the free scalar propagator.

Now, we calculate the path integral with respect to the transverse vector $%
x^{T}$ and $\psi \left( \tau \right) .$ The presence of a constant
electromagnetic field causes a particular quadraticity in the action of Eq. $%
(23)$, hence it is preferable to perform a fluctuation analysis on real
transverse bosonic vector $x^{T}$ and on fermionic vector $\psi ^{\mu }$ in
order to extract the contribution of the fixed action (classical action) in
the propagator, then%
\begin{eqnarray}
x^{T} &=&X^{T}+Y^{T} \\
x_{b,a}^{T} &=&X_{a,b}^{T}+Y_{a,b}^{T}
\end{eqnarray}%
and
\begin{equation}
\psi ^{\mu }\left( \tau \right) =\psi _{c}^{\mu }\left( \tau \right) +\zeta
^{\mu }\left( \tau \right) ,
\end{equation}%
where $X^{T}$ and $\zeta ^{\mu }\left( \tau \right) $ are respectively the
real and odd Grassmann fluctuations analysis. $\psi _{c}^{\mu }\left( \tau
\right) $ is fixed by the Euler-Lagrange equations (see Appendix A) and the
cyclic boundary conditions on the fluctuations $\zeta ^{\mu }\left( \tau
\right) $ is chosen as
\begin{equation}
E_{0}=\zeta ^{\mu }\left( 1\right) +\zeta ^{\mu }\left( 0\right) =0,
\end{equation}%
then the cyclic boundary condition of classical paths is

\begin{equation}
\psi _{c}^{\mu }\left( 1\right) +\psi _{c}^{\mu }\left( 0\right) =\Gamma
^{\mu }.
\end{equation}%
\

Let us consider all contributions given by Eqs. $\left( 24\right) $ and $%
\left( 26)-(30\right) $ for the evaluation of $G_{g}\left(
x_{b},x_{a}\right) $: it becomes%

\begin{align}
G_{g}(x_{b},x_{a})=&\notag
\frac{-i}{2}\exp(i\gamma^{\mu}\frac{\partial}{\partial\Gamma^{\mu}})
\int_{0}^{\infty}de_{0} \int\frac{dp^{L}}{(2\pi)^{2}}\int
d\varphi_{b}d\varphi_{a}d\eta_{b}d\eta_{a}\\\notag&\int D\eta
Dp_{\eta} \int_{E_{0}}\widetilde{D}\zeta^{T}\widetilde{D}\zeta^{L}
\int dp_{\sigma}\int_{Y_{a}}^{Y_{b}}DX^{T}
\delta(\varphi_{b}-\varphi_{a}+e_{0}k.p^{L})\delta(\varphi_{a}-k.x_{a})
\\\notag&\exp\{ip^{L}.(x_{b}^{L}-x_{a}^{L})+\frac{ie_{0}}{2}(p^{L2}-m^{2})
+i\int_{0}^{1}[-\frac{(\dot{X}^{T})^{2}}{2e_{0}}-\frac{g}{2}X^{T}.f\dot{X}^{T}]d\tau
\\\notag&-i\frac{g}{2}(\int_{\varphi_{a}}^{\varphi_{b}}d\varphi
A^{p}(\varphi)\frac{dY^{T}}{d\varphi}
+X^{T}.fY^{T}|_{\varphi_{a}}^{\varphi^{b}})+i\int_{0}^{1}[-ie_{0}g\eta
A'^{p}\psi^{T}_{c}
\\\notag&+p_{\eta}\dot{\eta}-ie_{0}g\zeta^{T}.f\zeta^{T}+i\zeta^{L}.\dot{\zeta}^{L}+i\zeta^{T}.\dot{\zeta}^{T}]d\tau
\\&+\psi_{c}(0).\psi_{c}(1)+p_{\sigma}(\eta_{a}-k(\psi_{ca}^{L}+\zeta^{L}(0)))\}|_{\Gamma=0},
\end{align}
by fixing the path $Y^{T}$ as
\begin{equation}
\left( -\frac{\dot{Y}^{T}}{e_{0}}+gfY^{T}-gA^{p}\left( \varphi \right)
\right) =0,
\end{equation}%
and using%
\begin{equation}
\int_{E}\tilde{D}\psi =\int_{E_{0}}\tilde{D}\zeta \text{ \ \ and \ \ \ }%
\int_{x_{a}^{T}}^{x_{b}^{T}}Dx^{T}=\int_{X_{a}}^{X_{b}}DX^{T}.
\end{equation}

The term which is a function of the real fluctuations $X^{T}$ appearing in
the action of expression $\left( 31\right) $ is equivalent to a known
gaussian integral related to the problem of scalar particle (without (SF))
in a constant electromagnetic field\ $\left[ 10\right] $. In fact, by using
the explicit definition of electromagnetic tensor $\left( 8\right) ,$ we can
show that this gaussian takes a particular form in terms of the uniform
constant magnetic field $B$ and the components of $X_{a}^{T}$ ,$X_{b}^{T}$\
(see Appendix B), where the components of $X_{a}^{T}$ ,$X_{b}^{T}$ in the
transverse plane are defined as%
\begin{equation}
X_{a}^{T}=\left(
\begin{array}{c}
X_{a}^{1} \\
X_{a}^{2}%
\end{array}%
\right) ,\text{ }X_{b}^{T}=\left(
\begin{array}{c}
X_{b}^{1} \\
X_{b}^{2}%
\end{array}%
\right) .
\end{equation}

The (GF) should be in a symmetrical form with respect to the initial and
final points in order to extract the wave functions. Therefore we symmetrize
the delta function $\delta \left( \varphi _{b}-\varphi
_{a}+e_{0}k.p^{L}\right) $ by inserting its exponential form
\begin{eqnarray}
\delta \left( \varphi _{b}-\varphi _{a}+e_{0}k.p^{L}\right)  \nonumber \\
=\int dz\exp \left[ iz\left( \varphi _{b}-\varphi
_{a}+e_{0}k.p^{L}\right) \right]
\end{eqnarray}%
into Eq. $\left( 31\right) $ and shifting the vector $p^{L}$ to $p^{L}-zk$ .
After integrating again over $z,$ we find%

\begin{align}
G_{g}(x_{b},x_{a})=&\notag
\frac{-i}{2}\exp(i\gamma^{\mu}\frac{\partial}{\partial\Gamma^{\mu}})
\int_{0}^{\infty}de_{0}
\int_{E_{0}}\widetilde{D}\zeta^{T}\widetilde{D}\zeta^{L}Dp_{\eta}
\\\notag&\int D\eta\int\frac{dp^{L}}{(2\pi)^{2}}\int
d\varphi_{b}d\varphi_{a}d\eta_{b}d\eta_{a}\int dp_{\sigma}
(\frac{igB}{4\pi\sin(\frac{e_{0}gB}{2})})\\\notag&\delta(\varphi_{a}-k.x_{a})\delta(\varphi_{b}-k.x_{b})
\exp\{ip^{L}.(x_{b}^{L}-x_{a}^{L})+\frac{ie_{0}}{2}(p^{L2}-m^{2})
\\\notag&-i\frac{g}{2}(\int_{\varphi_{a}}^{\varphi_{b}}d\varphi A^{p}(\varphi)\frac{dY^{T}}{d\varphi}
+X^{T}.fY^{T}|_{\varphi_{a}}^{\varphi^{b}})\}
\\\notag&\exp\{i\frac{gB}{2}[(X_{b}^{1}X_{a}^{2}-X_{b}^{2}X_{a}^{1})
-\frac{1}{2}\cot(\frac{e_{0}gB}{2})
\\\notag&((X_{b}^{1}-X_{a}^{1})^{2}+X_{b}^{2}-X_{a}^{2})^{2})]
+i\int_{0}^{1}[-ie_{0}g\eta A'^{p}\psi^{T}_{c}
+p_{\eta}\dot{\eta}-ie_{0}g\zeta^{T}.f\zeta^{T}
\\&+i\zeta^{L}.\dot{\zeta}^{L}+i\zeta^{T}.\dot{\zeta}^{T}]d\tau
+\psi_{c}(0).\psi_{c}(1)
+ip_{\sigma}(\eta_{a}-k.\psi_{ca}^{L})-ip_{\sigma}k.\zeta^{L}(0)\}|_{\Gamma=0}.
\end{align}

Notice that the part which does not contain the integration over Grassmann
paths in Eq. $\left( 36\right) $ describes the (GF) of a scalar particle in
both plane wave and constant magnetic fields $\left[ 14\right] .$

The only remaining path integral in the Eq. $(36)$, is the (SF). With the
help of the velocity variables $\omega _{\mu }\left( \tau \right) $ $\left[ 8%
\right] ,$ we compute the gaussian integrals with respect to the $\zeta ^{T}$
and $\zeta ^{L}$ (see Appendix B). After integrating successively over $%
p_{\eta }$ and $\eta $ in Eq. $(36),$ we deduce that the spin current
projected along the wave vector $k$ is constant during the evolution and
satisfies the equation%
\begin{equation}
\dot{\eta}=0,\text{ }\eta =\eta _{a}=\eta _{b}.
\end{equation}

Taking into account the classical equations in appendix A$,$ we deduce that
\begin{equation}
\text{\ }k.\dot{\psi}_{c}^{L}=0,\text{ \ }k.\psi _{c}^{L}\left( 1\right)
=k.\psi _{c}^{L}\left( 0\right) .
\end{equation}

The multiplication by the wave vector $k$ on the left of the boundary
condition $\left( 30\right) ,$ and the successive integrations over $%
p_{\sigma }$ and $\eta _{a}$ in Eq. $\left( 36\right) ,$ lead to
\begin{equation}
\eta _{a}=k.\psi _{c}^{L}\left( 0\right) =\frac{k.\Gamma ^{L}}{2}.
\end{equation}%
This equation preserves the induced condition by the projection of Eq. $%
\left( 4\right) $ along $k.$

It has been shown that all path integrals are reduced to the computed
gaussian integrals. What remains is the contribution of the effective
classical action in the calculation of the (GF). By substituting all
obtained solutions in appendix A into Eq. $\left( 39\right) $, and deriving
with respect to $\Gamma $, we find%

\begin{align}
G_{g}(x_{b},x_{a})=&\notag \frac{-i}{2}
\int_{0}^{\infty}de_{0}
\int\frac{dp^{L}}{(2\pi)^{2}}\int d\varphi_{b}d\varphi_{a}
(\frac{igB}{4\pi\sin(\frac{e_{0}gB}{2})})\delta(\varphi_{a}-k.x_{a})\delta(\varphi_{b}-k.x_{b})
\\\notag&\exp\{ip^{L}.(x_{b}^{L}-x_{a}^{L})+\frac{ie_{0}}{2}(p^{L2}-m^{2})-i\frac{g}{2}
(\int_{\varphi_{a}}^{\varphi_{b}}d\varphi
A^{p}(\varphi)\frac{dY^{T}}{d\varphi}
+X^{T}.fY^{T}|_{\varphi_{a}}^{\varphi^{b}})
\\\notag&+i\frac{gB}{2}[(X_{b}^{1}X_{a}^{2}-X_{b}^{2}X_{a}^{1})
-\frac{1}{2}\cot(\frac{e_{0}gB}{2})
((X_{b}^{1}-X_{a}^{1})^{2}+(X_{b}^{2}-X_{a}^{2})^{2})]\}
\\\notag&\{e^{(\frac{ie_{0}g}{2}B)}[1-\gamma^{\mu}\gamma^{\nu}k_{\mu}\epsilon^{*}_{\nu}K(\varphi_{b})]
\frac{\gamma^{\mu}\gamma^{\nu}\epsilon_{\mu}\epsilon^{*}_{\nu}}{2}[1+\gamma^{\mu}\gamma^{\nu}k_{\mu}\epsilon_{\nu}
K^{*}(\varphi_{a})]\\&+e^{-(\frac{ie_{0}g}{2}B)}[1-\gamma^{\mu}\gamma^{\nu}k_{\mu}\epsilon_{\nu}
K^{*}(\varphi_{b})]\frac{\gamma^{\mu}\gamma^{\nu}\epsilon^{*}_{\mu}\epsilon_{\nu}}{2}
[1+\gamma^{\mu}\gamma^{\nu}k_{\mu}\epsilon^{*}_{\nu}K(\varphi_{a})]\}
\end{align}
with%
\begin{eqnarray}
K\left( \varphi \right) &=&\frac{g}{2\left( k.p^{L}\right) }\exp \left[
\frac{igB\varphi }{\left( k.p^{L}\right) }\right]  \notag \\
&&\int_{\varphi _{0}}^{\varphi }d\varphi ^{\prime }\exp \left[ \frac{%
igB\varphi ^{\prime }}{\left( k.p^{L}\right) }\right] \left( \epsilon
.A^{\prime p}\right) .
\end{eqnarray}

Here we have used the formulas
\begin{eqnarray}
&&1-\frac{1}{2}\tanh \left( \frac{\alpha }{2}\right) \gamma ^{\mu }\gamma
^{\nu }\left( \epsilon _{\mu }\epsilon _{\nu }^{\ast }-\epsilon _{\mu
}^{\ast }\epsilon _{\nu }\right)   \notag \\
&=&\left( \frac{e^{-\frac{\alpha }{2}}}{e^{-\frac{\alpha }{2}}+e^{\frac{%
\alpha }{2}}}\right) \gamma ^{\mu }\gamma ^{\nu }\epsilon _{\mu }\epsilon
_{\nu }^{\ast }+\left( \frac{e^{\frac{\alpha }{2}}}{e^{-\frac{\alpha }{2}%
}+e^{\frac{\alpha }{2}}}\right) \gamma ^{\mu }\gamma ^{\nu }\epsilon _{\mu
}^{\ast }\epsilon _{\nu },
\end{eqnarray}%
and%
\begin{equation}
\gamma ^{\mu }\gamma ^{\nu }\left( \frac{\epsilon _{\mu }\epsilon _{\nu
}^{\ast }}{2}+\frac{\epsilon _{\mu }^{\ast }\epsilon _{\nu }}{2}\right) =1.
\end{equation}

$K^{\ast }\left( \varphi \right) $ is the conjugate of $K\left( \varphi
\right) .$The final expression of $G_{g}\left( x_{b},x_{a}\right) $ is quite
symmetric and is identical to the one obtained in $\left[ 15\right] .$%
\bigskip

It has been shown in [8] that the exact (SF) of a Dirac particle interacting
with a plane wave field can only be described through the classical
Grassmann paths due to the remarkable properties of the field, and
therefore, there is no contribution of the fluctuating trajectories.
Similarly, in our case, by either considering a weak magnetic field ($B\ll 1$%
) or neglecting the fluctuations around the classical paths in the second
order variation ($\zeta ^{\mu }\zeta ^{\nu }\approx 0$) as a semi-classical
calculation of the (SF), we find the gaussian integral given in appendix $%
(B-3)$ is the unity. However, the description of the spin interaction is
only presented by using the classical Grassmann trajectories without
performing any integration. In other words, the existence of a constant
magnetic field $B$ requires an introduction of fluctuating paths which
contribute as gaussian integral to the exact computation of the (SF).

By taking the limit case of the wave vector $k\rightarrow 0$ in Eq. (40), we
can deduce the influence of the constant magnetic field $B$ on the particle,
we, then, get%

\begin{align}
G_{g}(x_{b},x_{a})=&\notag \frac{-i}{2}
\int_{0}^{\infty}de_{0}
\int\frac{dp^{L}}{(2\pi)^{2}}
(\frac{igB}{4\pi\sin(\frac{e_{0}gB}{2})})
\exp\{ip^{L}.(x_{b}^{L}-x_{a}^{L})+\frac{ie_{0}}{2}(p^{L2}-m^{2})
\\\notag&+i\frac{gB}{2}[(X_{b}^{1}X_{a}^{2}-X_{b}^{2}X_{a}^{1})
-\frac{1}{2}\cot(\frac{e_{0}gB}{2})
((X_{b}^{1}-X_{a}^{1})^{2}+(X_{b}^{2}-X_{a}^{2})^{2})]\}
\\&\{e^{(\frac{ie_{0}g}{2}B)}
\frac{\gamma^{\mu}\gamma^{\nu}\epsilon_{\mu}\epsilon^{*}_{\nu}}{2}+e^{-(\frac{ie_{0}g}{2}B)}\frac{\gamma^{\mu}\gamma^{\nu}\epsilon^{*}_{\mu}\epsilon_{\nu}}{2}
\}
\end{align}

\section{\label{sec:level1 copy(3)}Conclusion}

We have calculated the exact Green function (GF) for a Dirac particle
interacting with a plane wave and constant transverse magnetic fields via
the formalism of Alexandrou et al. (global projection). In this approach,
the description of spin is established through the\ integration over
anticommuting Grassmann trajectories. In addition, the calculation of the
(GF) is based on two techniques, in the first one, we have introduced the
so-called constraints (functional identities) into the formulation. These
identities reduce the dimension of the integration of all projected paths
along the wave vector, and extract some classical trajectories in natural
way from the propagator. Indeed, it is shown that this method is very useful
for the path-integral approach particularly in the presence of the plane
wave field.

Due to the existence of a constant magnetic field $B$, we have adopted a
second technique which is based on a fluctuation analysis performed on real
and Grassmann variables. In fact, the paths are written in terms of both,
fixed and fluctuating trajectories, as a consequence, the path integral is
reduced to a calculation of known gaussian integrals, and by inserting the
classical solutions of Euler-Lagrange into the effective classical action,
we obtain the exact result of the (GF) as provided in the literature.

\bigskip \bigskip \appendix

\section{Classical solutions}

$\psi _{c}^{\mu }\left( \tau \right) $ is fixed by Euler-Lagrange equations
as
\begin{equation}
\dot{\psi}_{c}^{T}-e_{0}gf\psi _{c}^{T} =-e_{0}g\eta A^{\prime
p},\tag*{(A-1)}
\end{equation}
\begin{equation}
\dot{\psi}_{c}^{L}\left( \tau \right) =-\frac{i}{2}k\dot{p}_{\eta
}.\tag*{(A-2)}
\end{equation}

The classical solutions of $\psi _{c}^{T}\left( \tau \right) $ from the
Euler-Lagrange equations are%
\begin{equation}
\psi _{c}^{T}\left( \tau \right) =-e_{0}g\eta _{a}e^{Q\tau }\int_{0}^{\tau
}\left( e^{-Q\acute{\tau}}A^{\prime p}\right) d\acute{\tau}+e^{Q\tau }\psi
_{c}^{T}\left( 0\right)  \tag*{(A-3)}
\end{equation}

with
\begin{equation}
Q_{\mu \nu }=e_{0}gf_{\mu \nu }.  \tag*{(A-4)}
\end{equation}

We can obtain the initial and final classical solutions $\psi _{c}^{T}\left(
0\right) $ and $\psi _{c}^{T}\left( 1\right) $ of the spin variables from
the boundary condition $\left( 30\right) $ and the general solution (A-3)
\begin{align}
\psi _{c}^{T}\left( 0\right) =e_{0}g\eta _{a}e^{Q}\left(
1+e^{Q}\right) ^{-1}\int_{0}^{1}\left( e^{-Q\acute{\tau}}A^{\prime
p}\right) d\acute{\tau}
\notag & \\
+\frac{1}{2}\left( 1-\tanh \frac{Q}{2}\right) \Gamma ^{T},
\tag*{(A-5)}
\end{align}%
\begin{align}
\psi _{c}^{T}\left( 1\right) =-e_{0}g\eta _{a}e^{Q}\left(
1+e^{Q}\right) ^{-1}\int_{0}^{1}\left( e^{-Q\acute{\tau}}A^{\prime
p}\right) d\acute{\tau}
\notag &\\
+e^{Q}\left( 1+e^{Q}\right) ^{-1}\Gamma ^{T}.\tag*{(A-6)}
\end{align}
In the same way as for the transverse classical solution, we have
\begin{equation}
\psi _{c}^{L}\left( \tau \right) =-\frac{i}{2}kp_{\eta }+\frac{i}{2}kp_{\eta
_{a}}+\psi _{c}^{L}\left( 0\right)  \tag*{(A-7)}
\end{equation}%
and%
\begin{equation}
\psi _{c}^{L}\left( 1\right) =\frac{i}{4}k\left( p_{\eta
_{a}}-p_{\eta _{b}}\right) +\frac{\Gamma ^{L}}{2},\tag*{(A-8)}
\end{equation}
\begin{equation}
 \psi _{c}^{L}\left( 0\right)
=-\frac{i}{4}k\left( p_{\eta _{a}}-p_{\eta _{b}}\right)
+\frac{\Gamma ^{L}}{2}.\tag*{(A-9)}
\end{equation}

\section{Gaussian integrals}

The known gaussian integral that appears in the problem of scalar particle
(without (SF)) in constant electromagnetic field\ $\left[ 10\right] $ is

\begin{gather}
\int DX^{T}\exp \left[ i\int_{0}^{1}\left( -\frac{\left( \dot{X}^{T}\right)
%TCIMACRO{\U{b2}}%
%BeginExpansion
{{}^2}%
%EndExpansion
}{2e_{0}}-\frac{g}{2}X^{T}.f\dot{X}^{T}\right) d\tau \right]  \notag \\
=\frac{igB}{4\pi \sin \left( \frac{e_{0}gB}{2}\right) }\exp \left\{ i\frac{gB%
}{2}\left[ \left( X_{b}^{1}X_{a}^{2}-X_{b}^{2}X_{a}^{1}\right) \right.
\right.  \notag \\
\left. -\frac{1}{2}\cot \left( \frac{e_{0}g}{2}B\right) \left( \left(
X_{b}^{1}-X_{a}^{1}\right)
%TCIMACRO{\U{b2}}%
%BeginExpansion
{{}^2}%
%EndExpansion
+\left( X_{b}^{2}-X_{a}^{2}\right)
%TCIMACRO{\U{b2}}%
%BeginExpansion
{{}^2}%
%EndExpansion
\right) \right] .  \tag*{(B-1)}
\end{gather}

With the help of velocity variables $\omega _{\mu }\left( \tau \right) $ $%
\left[ 8\right] ,$ such that

\begin{align}
\omega _{\mu }\left( \tau \right) =\dot{\zeta}_{\mu }\left( \tau
\right) ,
\notag &\\
\zeta _{\mu }\left( \tau \right)
=\frac{1}{2}\int_{0}^{1}\varepsilon
\left( \tau -s\right) \omega _{\mu }\left( s\right) ds,  \notag &\\
\varepsilon \left( \tau \right) =\text{sign of }\tau ,\tag*{(B-2)}
\end{align}%
the integral of (SF) along $\zeta ^{T}$ reduce to a simple gaussian which
appears in the treatment of Dirac particle in a constant electromagnetic
field
\begin{align}
\int_{E_{0}}\tilde{D}\zeta ^{T}\exp \left\{ i\int_{0}^{1}\left[
-ie_{0}g\zeta ^{T}.f\zeta ^{T}+i\zeta ^{T}.\dot{\zeta}^{T}\right]
d\tau
\right\}  \notag &\\
=\left[ \det \left( \cosh \left( \frac{e_{0}g}{2}f\right) \right)
\right] ^{1/2}=\cosh \left( \frac{ie_{0}gB}{2}\right)
,\tag*{(B-3)}
\end{align}%
thus the integral according to the longitudinal components
\begin{equation}
\int_{E_{0}}\tilde{D}\zeta ^{L}\exp \left\{ i\int_{0}^{1}\left( i\zeta ^{L}.%
\dot{\zeta}^{L}-p_{\sigma }k\zeta ^{L}\left( 0\right) \right) \right\} =1,
\tag*{(B-4)}
\end{equation}%
since it is of type%
\begin{gather}
\int \tilde{D}\omega ^{L}\exp \left[ \int \left( \omega ^{L}\left( \tau
\right) \varepsilon \left( \tau -s\right) \omega ^{L}\left( s\right) \right.
\right.  \notag \\
\left. \left. +I\omega ^{L}\left( s\right) \right) d\tau ds\right] =1,
\tag*{(B-5)}
\end{gather}%
with%
\begin{equation}
I_{\mu }=\frac{1}{2}k_{\mu }p_{\sigma },\tag*{(B-6)}
\end{equation}%
where $I_{\mu }$ does not depend on time of evolution and $I^{2}=0.$

\end{document}